\documentclass{Interspeech}
\usepackage{tablefootnote}
\usepackage{multirow}
\interspeechcameraready

\title{Pushing the Performance of Synthetic Speech Detection with Kolmogorov-Arnold Networks and Self-Supervised Learning Models}

\author[affiliation={1,2}]{Tuan Dat}{Phuong$^*$}
\author[affiliation={1,2}]{Long-Vu}{Hoang$^*$}
\author[affiliation={2}]{Huy Dat}{Tran}

\affiliation{SoICT}{Hanoi University of Science and Technology}{Vietnam}
\affiliation{Institute for Infocomm Research (I$^2$R)}{A$^*$STAR}{Singapore}
\email{\{phuongtuandat2915, longvu200502\}@gmail.com, hdtran@i2r.a-star.edu.sg}

\usepackage{comment}

\begin{document}

\maketitle

\begin{abstract}
    
    %Recent advancements in speech synthesis technologies have led to increasingly sophisticated spoofing attacks, posing significant challenges for automatic speaker verification systems. While systems based on self-supervised learning (SSL) models, particularly the XLSR-Conformer architecture, have demonstrated remarkable performance in synthetic speech detection, there remains room for architectural improvements. In this paper, we propose a novel approach that replaces the traditional Multi-Layer Perceptron (MLP) in the XLSR-Conformer model with a Kolmogorov-Arnold Network (KAN), a powerful universal approximator based on the Kolmogorov-Arnold representation theorem. Our experimental results on ASVspoof2021 demonstrate that the integration of KAN to XLSR-Conformer model can improve the performance by 60.55\% relatively in Equal Error Rate (EER) LA and DF sets, further achieving 0.70\% EER on the 21LA set. Besides, the proposed replacement is also robust to various SSL architectures. These findings suggest that incorporating KAN into SSL-based models is a promising direction for advances in synthetic speech detection.

    Recent advancements in speech synthesis technologies have led to increasingly advanced spoofing attacks, posing significant challenges for automatic speaker verification systems. While systems based on self-supervised learning (SSL) models, particularly the XLSR-Conformer model, have demonstrated remarkable performance in synthetic speech detection, there remains room for architectural improvements. In this paper, we propose a novel approach that replaces the traditional Multi-Layer Perceptron in the XLSR-Conformer model with a Kolmogorov-Arnold Network (KAN), a novel architecture based on the Kolmogorov-Arnold representation theorem. Our results on ASVspoof2021 demonstrate that integrating KAN into the SSL-based models can improve the performance by 60.55\% relatively on LA and DF sets, further achieving 0.70\% EER on the 21LA set. These findings suggest that incorporating KAN into SSL-based models is a promising direction for advances in synthetic speech detection.\let\thefootnote\relax\footnotetext{$^*$Equal contribution. Work done during their internship at A$^*$STAR, Singapore.}

    % The latest state-of-the-art models in synthetic speech detection leverage rich representation from self-supervised learning (SSL) models, project to lower dimensions with a linear transformation, and feed to a powerful sequence-to-sequence architecture like Conformer.
\end{abstract}

\section{Introduction}
\label{sec:intro}
In recent years, speech synthesis technologies have achieved remarkable progress, enabling the generation of increasingly more natural and convincing synthetic voices. While these advancements in text-to-speech (TTS) and voice conversion (VC) systems demonstrate the potential of conversational AI applications in human-computer interaction, they also raise significant security concerns in biometric authentication systems, particularly in speaker verification applications \cite{10.1109/TASLP.2020.3038524, 10.1145/3552466.3556531, hoang24b_interspeech}. Therefore, it is crucial to develop effective and robust systems for synthetic speech detection (SSD) as a means of protection \cite{https://doi.org/10.48550/arxiv.2210.00417}.

Recent state-of-the-art SSD systems leverage Self-supervised learning (SSL) as input features to sequence-to-sequence architectures for synthetic speech classification. SSL models can learn powerful representations by being trained on a large amount of unlabeled data. As a result, SSL models can be applied very effectively to various downstream tasks (e.g., speaker verification, emotion recognition, automatic speech recognition, etc.) \cite{rosello23_interspeech, truong24b_interspeech, vaessen2022fine, chen2023exploring, 9979878, gupta22_interspeech, liu2025nes2net, 10889972} when fine-tuned on a limited amount of labelled data. In synthetic speech detection, SSL models can be used as feature extractors, followed by a projection with Multi-Layer Perceptrons (MLPs) and a combination with other architectures such as Conformer \cite{gulati2020conformerconvolutionaugmentedtransformerspeech} or AASIST \cite{jung2021aasistaudioantispoofingusing} for SSD and deepfake detection tasks \cite{rosello23_interspeech, tak2022automaticspeakerverificationspoofing}.

For a long time, MLPs have been an indispensable component in deep learning models, particularly as fundamental architectures for models like Transformers \cite{10096278} or Convolutional Neural Networks (CNNs) \cite{inproceedings}. 
However, MLPs are often challenged by high dimensional data like images \cite{lecun1998gradient} owing to their lack of capacity to inherently capture spatial patterns in the data, resulting in scalability and efficiency concerns and may lead to suboptimal performance. Given that speech signals are mostly represented in raw waveforms, spectrograms or deep embeddings of high dimensionalities, the incorporation of MLP may result in suboptimal performance.

The recent emergence of KANs \cite{liu2024kankolmogorovarnoldnetworks}, a class of neural networks inspired by the Kolmogorov-Arnold representation theorem, introduces an innovation to the neural networks by employing learnable activation functions. The Kolmogorov-Arnold representation theorem assumes that any multivariate continuous function can be expressed as a sum of continuous functions of a single variable. Based on that, KANs replace the fixed activation functions in MLPs with learnable univariate functions, enabling a more flexible and interpretable framework for function approximation \cite{SCHMIDTHIEBER2021119}. Recent works adapting KANs also claimed that this new architecture could address various intrinsic limitations of MLPs, especially in handling complex functional mapping in high-dimensional spaces \cite{liu2024kankolmogorovarnoldnetworks, azam2024suitability, li2024investigation}, particularly promising for speech data.
%This structural difference sets KANs apart from MLPs, as it allows for greater adaptability and better alignment with the decomposition of multivariate functions. 
% The growing interest in KANs stems from their ability to achieve comparable or even superior accuracy to larger MLPs, while also exhibiting faster neural scaling laws and enhanced interpretability \cite{somvanshi2024surveykolmogorovarnoldnetwork}. These properties make KANs a promising alternative to traditional MLP-based architectures in a variety of deep learning applications.

% Building upon the success of KANs, the Group Rational KAN (GR-KAN) \cite{yang2024kolmogorovarnoldtransformer} architecture has been introduced as a refinement to the standard KAN model. The key innovation in GR-KAN is the replacement of the traditional B-spline basis functions with rational functions. This change allows for the sharing of function coefficients and base functions among groups of edges, leading to a more efficient and stable model. The use of rational functions in GR-KANs addresses some of the limitations inherent in standard KANs, such as the computational cost and the potential instability during the training process. By leveraging the advantages of rational functions, GR-KANs provide a more robust and optimized approach to function approximation, making them a promising candidate for various machine learning tasks, including speech anti-spoofing and deepfake detection.

In this paper, we propose a novel system enhanced with KAN architecture to bridge the gap on various SSD benchmarks. We employ the pre-trained XLS-R model and Conformer encoders, enhanced with KAN architecture to approximate the high-dimensional features for a better capability of detecting artifacts in synthetic speech. Our model achieves the new state-of-the-art (SOTA) results of 0.80\% and 0.70\% on the ASVspoof2021 LA set under fixed- and variable-length conditions respectively, while remains on par with other competitive systems. The subsequent sections in this paper are organized as follows: Section \ref{sec:theory} presents the theoretical formulation of KANs and their extension, Section \ref{sec:method} describes the methods employed in this paper, and Section \ref{sec:experiments} discusses the experimental implementations and results to compare our model with other SOTA models. Section \ref{sec:conclusion} concludes our contributions in this paper.

\section{Theoretical Formulation}
\label{sec:theory}

% The following section outlines all methods utilized in this research, including KANs, GR-KAN, the baseline model architecture, and our proposed new model architecture.

\subsection{Multi-Layer Perceptrons (MLPs)}
A Multi-Layer Perceptron (MLP) is a fully connected feedforward neural network consisting of multiple layers of neurons. Each neuron in a layer is connected to every node in the following layer, and the node applies a nonlinear activation function to the weighted sum of its inputs.

The foundation of MLP is supported by the Universal Approximation Theorem \cite{hornik1989multilayer}. The theorem states that a feedforward network with a single hidden layer containing a finite number of neurons can approximate any continuous function on compact subsets of $\mathbb{R}^n$, given appropriate activation functions.

\subsection{Kolmogorov-Arnold networks (KANs)}
\label{ssec:kan}

The Kolmogorov-Arnold representation theorem states that any continuous multivariate function can be represented as a sum of univariate functions. More specifically, given \textbf{x} a vector of dimension $n$, \(f\) a multivariate continuous function such that: $f : [0,1]^n \rightarrow\mathbb{R}$, it can be expressed that:

\begin{equation}
\label{eq:theorem}
    f(\textbf{x}) = \sum_{q=1}^{2n+1} \Phi_{q} (\sum_{p=1}^{n} \phi_{q,p}(x_{p})),
\end{equation}

where $\phi_{q,p}: [0,1] \rightarrow \mathbb{R}$ and $\Phi_{q} : \mathbb{R} \rightarrow\mathbb{R}$ are  univariate functions. Equation \ref{eq:theorem} demonstrates that the representation and computation of complex functions can be significantly simplified. This characteristic has motivated neural network topologies, particularly in the domains of function approximation and dimensionality reduction, and offers theoretical support for high-dimensional data modelling.

Leveraging this theorem, \cite{liu2024kankolmogorovarnoldnetworks} introduces a generalized KANs layer to learn activation functions, which are univariate functions on edge. Officially, a KANs layer with \(d_{in}\)-dimensional inputs and \(d_{out}\)-dimensional outputs is described in Equation \ref{eq:kan}.
\begin{equation}
\label{eq:kan}
    f(\textbf{x}) = \Phi \circ x = [\sum_{i=1}^{d_{in}} \phi_{i,1}(x_{i}) ,. . .  
  ,\sum_{i=1}^{d_{in}} \phi_{i,d_{out}}(x_{i})],
\end{equation}
where \(\phi_{i,j}\) is a univariate transformation, \(\Phi\) captures the combined mapping across the input dimensions. 

In KANs, each learnable univariate function \(\phi_{q,p}\) can be defined as a B-spline:

\begin{equation}
\label{eq:b-splines}
    \text{spline}(x) = \sum_{i} c_{i}B_{i}(x).
\end{equation} 

The activation function \(\phi(x)\) is a weighted combination of a basis function \(b(x)\) with the B-spline. Given \(\text{SiLU}(x) = \frac{x}{1 + e^{-x}}\), $w_b$ and $w_s$ are corresponding weights of the basis function and spline function, the activation function can be expressed as: 

\begin{equation}
\label{eq:phi}
    \phi(x) = w_{b}\cdot\text{SiLU}(x) + w_{s}\cdot\text{spline}(x),
\end{equation} 

Equation \ref{eq:overall}{} illustrates the general architecture of an $L$-layer KAN:
\begin{equation}
\label{eq:overall}
    \text{KAN(x)} = (\Phi_{L-1} \circ \Phi_{L-2} \circ ... \circ \Phi_{0})(x).
\end{equation} 

\subsection{Group Rational Kolmogorov-Arnold networks (GR-KANs)}
\label{ssec:grkan}

Yang et al. \cite{yang2024kolmogorovarnoldtransformer} suggests replacing B-spline with a rational function due to the limitations of standard KANs. GR-KAN divides the \(I\) input channels into \(k\) groups and shares the parameters of the rational functions across all channels within the same group. Equation \ref{eq:grkan_layer} describes the formula of a GR-KAN layer applied with input vector \(x\), where \(\phi\) is now a rational function, \(I/k\) represents the dimension size of each group, \(O\) is the dimension of the output vector, and \(w\) refers to unique scalars.

\begin{equation}
\label{eq:grkan_layer}
\begin{aligned}
L(\mathbf{x}) &= \Phi \circ \mathbf{x} \\
&= 
\begin{bmatrix}
\sum_{i=1}^{I} w_{i,1} \phi_{\lfloor \frac{i}{I/k} \rfloor}(x_i) & \cdots & \sum_{i=1}^{I} w_{i,O} \phi_{\lfloor \frac{i}{I/k} \rfloor}(x_i)
\end{bmatrix}
\end{aligned}
\end{equation}

The key improvement that GR-KAN introduces over KAN lies in its weight initialization strategy, which aims to ensure a variance-preserving effect across the network. This initialization helps prevent the increase or decrease in activation magnitudes throughout the layers, thereby promoting network stability. Additionally, GR-KAN weights can be seamlessly loaded with the weights from an MLP's linear layer, as the GR-KAN layer integrates both a linear layer and a group-wise rational layer.

\section{Proposed Methodology}
\label{sec:method}
\subsection{Baseline Model Architectures}
\label{ssec:xlsr}
\begin{figure}[h]
    \centering
    \includegraphics[width=0.5\textwidth]{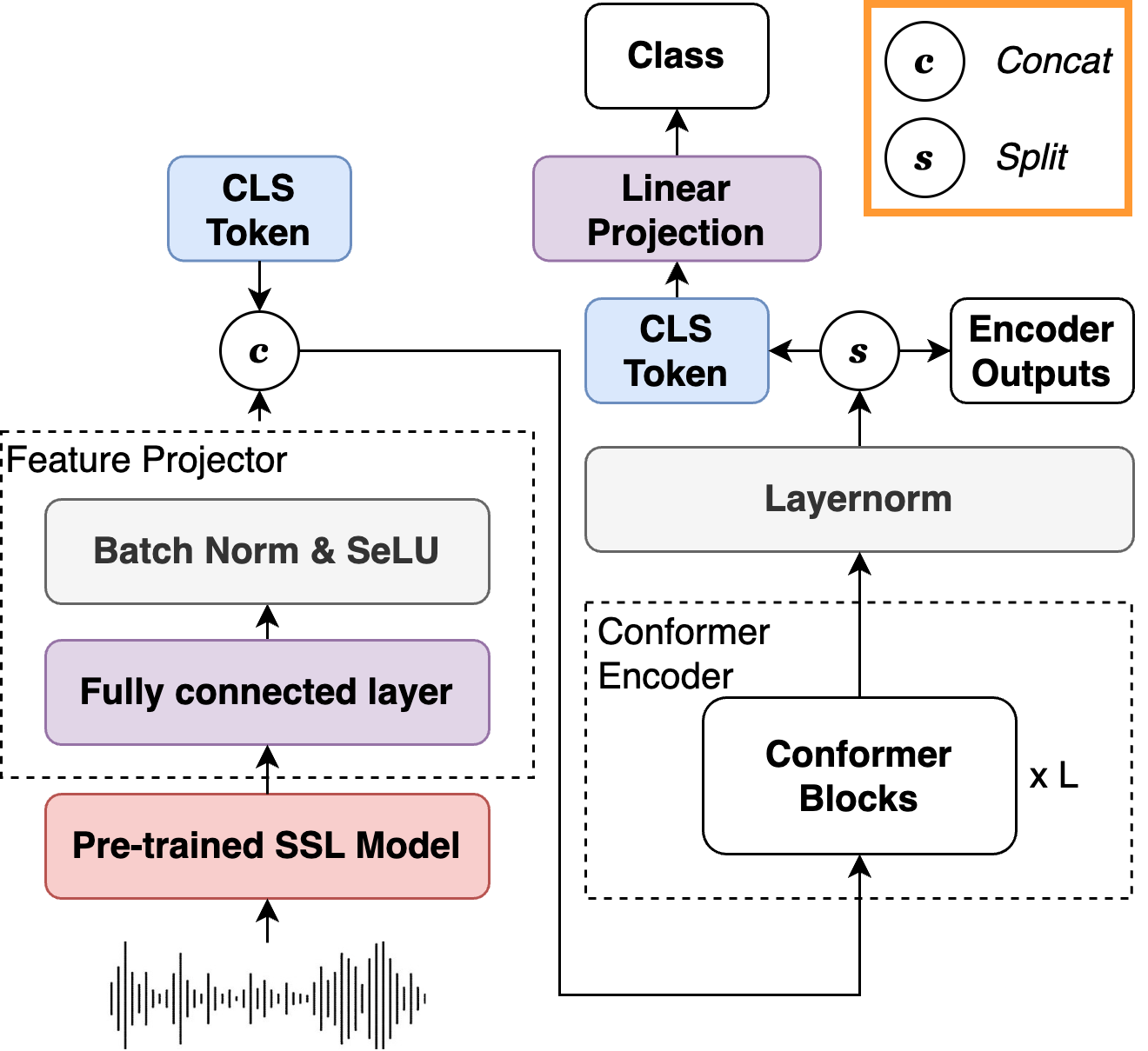}
    \caption{Architecture of the baseline XLSR-Conformer model. The XLSR-Conformer+TCM baseline only modifies the MHSA module.}
    \label{fig:baseline}
\end{figure}

We use two state-of-the-art architectures as our baseline, XLSR-Conformer \cite{truong24b_interspeech} and its variant XLSR-Conformer+TCM with an additional temporal-channel dependency modelling (TCM) module. As can be seen in Figure \ref{fig:baseline}, the XLSR-Conformer baseline comprises two main parts: (i) the pre-trained XLS-R \cite{babu2021xlsrselfsupervisedcrosslingualspeech}, which is a variant of the wav2vec 2.0 \cite{baevski2020wav2vec20frameworkselfsupervised} model, utilised as the feature extractor to capture contextualised representations from the high-dimensional speech signal; and (ii) the Conformer Encoder. Given an input speech signal $O$, the $T$-length output SSL features are denoted as $X = \text{SSL}(O) = (x_t \in \mathbb{R}^D | t = 1,..., T)$, with $D$ being the output dimension of the SSL model. 

The extracted features $X$ are projected to a lower dimension by an MLP with SeLU activation function before being fed to the Conformer Encoder. Out projected features are denoted as $\tilde{X} = \text{SeLU}(\text{Linear}(X))$, where $\tilde{X}=(\tilde{x}_t \in \mathbb{R}^{D'} | t=1,..., T)$. The Conformer Encoder is a stack of $L$ Conformer blocks, each containing a Multi-Head Self-Attention (MHSA) and a Convolutional Module sandwiched between two feed-forward modules. In order to adapt the sequence-to-sequence Conformer architecture for a solving classification task, a learnable classification token is prepended to the input embedding of the Conformer Encoder. $\tilde{X}_{in} = [\tilde{X}, \tilde{X}_\text{CLS}], \tilde{X} \in \mathbb{R}^{T \times {D'}}, \tilde{X}_{CLS} \in \mathbb{R}^{1 \times {D'}}$ is the input to the Conformer Encoder, where $[\cdot, \cdot]$ denotes the concatenation operation. Finally, the state of the classification token $\tilde{X}_{CLS}$ at the output of the last Conformer block is fed to a linear layer to classify the input speech signal as bonafide or spoof. During training, the classification token $\tilde{X}_{CLS}$ learns to capture the most relevant captures to distinguish a synthetic speech from a genuine one.

Truong et al. \cite{truong24b_interspeech} proposed integrating the channel representation head token into the temporal input token within the MHSA, called XLSR-Conformer+TCM model. Therefore, the model is capable of learning the temporal-channel dependencies from the input sequence. The architecture is similar to the XLSR-Conformer baseline, with the modifications only represented in the MHSA module. The XLSR-Conformer+TCM reduces the deepfake detection error rate by 26\% relatively while being competitive on the logical access benchmark of ASVspoof 2021.

In the aforementioned baseline architectures, we assume that the projection done by MLP may lead to suboptimal performance owing to the complexities in the high-dimensional contextualised representations. Therefore, we propose to replace the MLP with a simple GR-KAN layer to effectively approximate the functional mapping without an excessive computation overhead.

%With the help of the large-scale architecture and SSL-enabled training on vast amounts of data, SSL models, such as XLSR, are able to extract rich speech representations that are helpful for a variety of speech tasks.

% In additionally, our baseline model employ TCM to robust the ability of MHSA for capturing temporal-channel dependencies. As described in \cite{truong24b_interspeech}, the TCM architecture includes three components: Head Token Generation, Multi-Head Self-Attention, and Classification Token Enrichment. The TCM aims to create the head token for channel information and then incorporate the temporal-channel dependency into the initial temporal tokens to improve model's capability in SSD task.

\subsection{XLSR-GRKAN-Conformer Model}
\label{ssec:xlsr-kan}

\begin{figure}[h]
    \centering
    \includegraphics[width=0.17\textwidth]{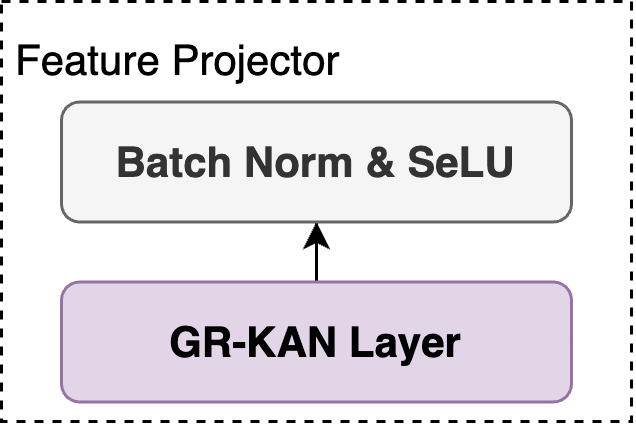}
    \caption{Architecture of feature projector with GR-KAN.}
    \label{fig:grkan}
\end{figure}

Our proposed architecture\footnote{Implementations at: https://github.com/HuSTeP-Human-Speech-Text-Processing-Lab/XLSR-GRKAN-Conformer} with XLSR-Conformer enhanced by GR-KAN is illustrated in Figure \ref{fig:grkan}. Given the demonstrated superiority of KAN in function approximation and dimensionality reduction compared to traditional MLP architectures, along with the proven ability of GR-KAN in maintaining training stability as discussed in Sections \ref{ssec:kan} and \ref{ssec:grkan}, we propose to replace the conventional MLP projection from XLS-R features to Conformer Encoder with the GR-KAN implementation. Therefore, the input to the Conformer Encoder is now represented by $\tilde{X}_{in} = [\tilde{X}_{\text{GRKAN}}, \tilde{X}_{CLS}] \in \mathbb{R}^{T \times {D'}}$, where  $\tilde{X}_{\text{GRKAN}} = \text{SeLU}(\text{GR-KAN}(X))$ and $X \in \mathbb{R}^{T \times D}$ is the same feature from an SSL model in Section \ref{ssec:xlsr}, $\text{GR-KAN}(\cdot)$ is determined by Equation \ref{eq:grkan_layer}.

This architectural modification is primarily motivated by the inherent requirement to reduce the dimensionality of speech representations derived from SSL models before they can serve as input to the sequence-to-sequence models like Conformer - a scenario where KAN's strengths become particularly advantageous. Furthermore, the GR-KAN's inherent capability to mitigate the problem of activation magnitude fluctuations across network layers potentially minimizes the loss of valuable features extracted by the SSL model. These architectural improvements collectively enhance the performance of the baseline model, as will be demonstrated in the following sections.

\section{Experiments}
\label{sec:experiments}
\begin{table*}[h]
\centering
\renewcommand{\arraystretch}{1.1}
\caption{Performance comparison with SOTA models on the ASVspoof 2021 LA and DF evaluation set using fixed-length (Fix) and variable-length (Var) utterance evaluation. The best result is bolded, dash denotes the results are unavailable, $\dagger$ denotes reproduced results.}
\label{tab:table1}
\begin{tabular}{l|cccc|cc}
\hline
{\textbf{Model}} & \multicolumn{2}{c}{\textbf{21LA (Fix)}} & \multicolumn{2}{c|}{\textbf{21LA (Var)}} & \textbf{21DF (Fix)} & \textbf{21DF (Var)}\\ \cline{2-7} 
 & \textbf{EER (\%)} & \textbf{min t-DCF} & \textbf{EER (\%)} & \textbf{min t-DCF} & \textbf{EER (\%)} & \textbf{EER (\%)}\\
 \hline
RawNet2 \cite{9414234} & 9.50 & 0.4257 & - & - & 22.38 & -\\ 
AASIST \cite{jung2021aasistaudioantispoofingusing} & 5.59 & 0.3398 & - & - & - & -\\ 
RawFormer \cite{10096278} & 4.98 & 0.3186 & 4.53 & 0.3088 & - & - \\ 
XLSR-AASIST \cite{tak2022automaticspeakerverificationspoofing} & 1.00 & 0.2120 & - & - & 3.69 & -\\ 
\hline
XLSR-Conformer \cite{rosello23_interspeech} (Baseline 1)$\dagger$  & 1.07 & 0.2136  & - & - & 2.55 & -\\ 

XLSR-Conformer + TCM \cite{truong24b_interspeech} (Baseline 2)$\dagger$ & 1.74 & 0.2333 & 2.13 & 0.2445 & 2.74 & 2.77\\ 

XLSR-GRKAN-Conformer (Proposed) & 1.05 & 0.2140  & 0.88  & 0.2085  & \textbf{1.95}  & \textbf{2.31} \\ 
XLSR-GRKAN-Conformer + TCM (Proposed) & \textbf{0.80} & \textbf{0.2067} & \textbf{0.70} & \textbf{0.2042} & 2.54 & 2.62\\ 
\hline

\end{tabular}
\end{table*}
\subsection{Datasets and Evaluation metrics}
\label{ssec:dataset}

The training and development datasets are sourced from the ASVspoof 2019 \cite{Todisco2019ASVspoof2F} logical access (LA) track, including bonafide speech and synthetic speech generated from two speech synthesis techniques: voice conversion and text-to-speech. We evaluate our model on the ASVspoof 2021 \cite{10155166} logical access (LA) and deepfake (DF) corpus. Two known and eleven unknown attack types are included in the ASVspoof 2021 LA evaluation set. To replicate real-world scenarios, speech data is subjected to various codec and compression changes. Additionally, in contrast to the LA set, ASVspoof 2021 presented a new deepfake (DF) evaluation set with two more source data sets. Our primary evaluation metrics
are the common-used equal error rate (EER) and minimum
normalized tandem detection cost function (min t-DCF).

% Besides, we also evaluated our baseline and proposed models on the most recent benchmark in synthetic speech detection to assess the performance on realistic unseen data. The additional evaluation sets include: (i) the In-the-wild \cite{} comprising 20.7 hours of bonafide and 17.2 hours of spoofed audio collected in the wild; and (ii) the hidden subset of ASVspoof 2021 with non-speech segments removed from the utterances, to evaluate the dependence of synthetic speech detection models upon certain data characteristics.

\subsection{Experimental Setup}
\label{ssec:setup}

To ensure a fair comparison, we retained all configurations mentioned in baseline models \cite{rosello23_interspeech, truong24b_interspeech}. During training, the audio data are either trimmed or padded to approximately 4 seconds. We used the Adam optimizer with a learning rate of $10^{-6}$ and a weight decay of $10^{-4}$. The final results are reported by averaging the top 5 best models on the validation set. Early stopping is applied if the validation loss does not improve after 7 epochs. We used the codebase in the baseline model and the official implementation of GR-KAN\footnote{https://github.com/Adamdad/kat.git}.

Additionally, we applied the RawBoost data augmentation technique to the training data. The configuration and parameters of RawBoost \cite{9746213} used in our experiment follow those in our baseline paper \cite{truong24b_interspeech}. For evaluation on the LA and DF tracks, we trained two distinct SSD systems with separate RawBoost settings: combining linear and nonlinear convolutive noise with impulsive, signal-dependent additive noise strategies for the LA track, and stationary, signal-independent additive, randomly coloured noise for the DF track.

% \subsection{Evaluation Metrics}

\subsection{Experimental Results}
\label{ssec:result}

% The performance of the proposed model with our reproduced baseline model XLSR-Conformer with TCM and various existing competitive models on the ASVspoof2021 LA and DF evaluation dataset is described in Table \ref{tab:table1}. For fixed-length input evaluation on the LA track, our model achieves an approximately 54\% relative improvement in EER compared to the baseline XLSR-Conformer with TCM (0.80\% vs 1.74\%). The integration of GR-KAN with the XLSR-Conformer not only establishes a new state-of-the-art performance with EERs of 0.80\% and 0.70\% on the ASVspoof LA track but also demonstrates superior performance over other models on the DF track. Notably, our model surpasses the reproduced baseline results by 7.3\% for fixed-length input and 5.42\% for variable-length input.

Table \ref{tab:table1} represents the results on ASVspoof21 LA and DF evaluation sets. Our proposed models, XLSR-GRKAN-Conformer and XLSR-GRKAN-Conformer + TCM, consistently outperform their corresponding baselines across all evaluation settings. Compared to XLSR-Conformer (Baseline 1), XLSR-GRKAN-Conformer achieves a relative improvement of 1.87\% in EER for 21LA (Fix) (1.07 $\rightarrow$ 1.05) and 17.8\% for 21LA (Var) (1.07 $\rightarrow$ 0.88). Similarly, the min t-DCF score improves from 0.2136 to 0.2085, reflecting better calibration. Against XLSR-Conformer + TCM (Baseline 2), our XLSR-GRKAN-Conformer + TCM model yields even stronger results, reducing EER by 54.0\% for 21LA (Fix) and 67.1\% for 21LA (Var), with a corresponding min t-DCF improvement of 12.7\%.

For the 21DF task, XLSR-GRKAN-Conformer outperforms Baseline 1 in both fixed-length (2.55 $\rightarrow$ 1.95, a 23.5\% relative reduction) and variable-length (2.55 $\rightarrow$ 2.31, a 9.4\% relative reduction) settings. Our XLSR-GRKAN-Conformer + TCM model also shows consistent improvement over Baseline 2, achieving a 7.3\% relative reduction in EER for 21DF (Fix) and 5.4\% for 21DF (Var). These results confirm that integrating GR-KAN into the Conformer-based system significantly enhances performance over the simple MLP, and the combination with TCM further amplifies these benefits.

\begin{table}[h]
\centering
\renewcommand{\arraystretch}{1.3}
\caption{Performance comparison with baseline models using variable-length data for both training and evaluation}
\label{tab:table2}
\begin{tabular}{l|c c|c}
\hline
{\textbf{System}} & \multicolumn{2}{c|}{\textbf{21LA}} & \textbf{21DF} \\ \cline{2-4} 
 & \textbf{EER (\%)} & \textbf{min t-DCF} & \textbf{EER (\%)} \\ \hline
Baseline 2 \cite{truong24b_interspeech} & 1.96 & 0.2387 & 2.56 \\ \hline
Our model & \textbf{0.79} & \textbf{0.2061} & \textbf{2.54} \\ \hline
\end{tabular}
\end{table}

Table \ref{tab:table2} presents a comparison between our model and the baseline model on both LA and DF tracks, using variable length data for both training and evaluation. Specifically, our model achieves a 59.7\% relative reduction in EER for 21LA and a 13.7\% improvement in min t-DCF (0.2387 → 0.2061), while maintaining comparable performance on 21DF (2.56\% $\rightarrow$ 2.54\%).

% Our method consistently outperforms the baseline model across various evaluation conditions, demonstrating that KANs can be effectively applied to speech processing tasks, particularly in the context of SSD tasks.

\subsection{Ablation Study and Analysis}
\label{sec:ablation}

\begin{table}[h!]
\centering
\caption{EER (\%) result with different SSL models}
\label{tab:ssl-ablations}
\begin{tabular}{l|lrr}
\hline
\textbf{SSL Model}                       & \textbf{Projector} & \multicolumn{1}{c}{\textbf{21LA (Fix)}} & \textbf{21DF (Fix)} \\ \hline
\multirow{2}{*}{WavLM Large\tablefootnote{https://huggingface.co/microsoft/wavlm-large}}                   & MLP                 & 4.07                                       & 11.25                   \\
                                         & GR-KAN              & 2.79                                    & 6.21                \\
\multirow{2}{*}{XLS-R-300M\tablefootnote{https://github.com/facebookresearch/fairseq}}                    & MLP                 & 1.07                                    & 2.55                \\
                                         & GR-KAN              & 1.05                                    & 1.95                \\
\multirow{2}{*}{UniSpeech-SAT\tablefootnote{https://huggingface.co/microsoft/unispeech-sat-base-plus}} & MLP                 & 17.46                                   & 28.46               \\
                                         & GR-KAN              & 14.78                                   & 15.58               \\
\multirow{2}{*}{mHuBERT-147\tablefootnote{https://huggingface.co/utter-project/mHuBERT-147}}             & MLP                 & 21.20                                   & 16.48               \\
                                         & GR-KAN              & 16.05                                   & 13.80               \\ \hline
\end{tabular}%

\end{table}
In this section, we assess the robustness of our proposed replacement of MLP with GR-KAN in working with features from various different SSL models of various sizes and architectures. We consider: (i) WavLM, a self-supervised model optimized for speech processing tasks; (ii) XLS-R, a cross-lingual variant of wav2vec 2.0 designed for multilingual speech representation; (iii) UniSpeech-SAT, which incorporates speaker-aware training to enhance speaker and content modeling; and (iv) mHuBERT-147, a multilingual version of HuBERT trained on 147 languages for robust speech representation. For XLS-R, we use the \texttt{fairseq} implementation, other SSL models employ \texttt{HuggingFace} implementations.

% \subsubsection{Position of GR-KAN}

% \usepackage{multirow}
Table \ref{tab:ssl-ablations} demonstrates the robustness of replacing MLP with GR-KAN across different SSL models. On average, replacing MLP with GR-KAN results in a 29.1\% relative reduction in EER across different SSL models and tasks, including WavLM, XLS-R, UniSpeech-SAT, and mHuBERT-147. The performance gains are particularly notable for high-EER models like UniSpeech-SAT and mHuBERT-147, where GR-KAN significantly enhances detection performance. These results confirm that GR-KAN is a more effective and generalizable alternative to MLP, making it a robust choice for various self-supervised learning features.

% Meanwhile, Table \ref{tab:table3} compares the performance of both models using Conformer encoders of different sizes. It is evident that regardless of the Conformer type, our model consistently outperforms the baseline \cite{truong24b_interspeech}, further reinforcing the effectiveness of GR-KAN compared to MLP.

% \begin{table}[h]
% \centering
% \renewcommand{\arraystretch}{1.3}
% \caption{Performance comparison with baseline models using different numbers of attention heads and embedding sizes, where ``-B'' corresponds to the Conformer encoder's size of 256 with 4 attention heads, and ``-L'' denotes 512 with 8 attention heads.}
% \label{tab:table3}
% \begin{tabular}{l|c c|c}
% \hline
% {\textbf{System}} & \multicolumn{2}{c|}{\textbf{21LA}} & \textbf{21DF} \\ \cline{2-4} 
%  & \textbf{EER (\%)} & \textbf{min t-DCF} & \textbf{EER (\%)} \\ \hline
% Baseline 2 \cite{truong24b_interspeech} -B & 1.28 & 0.2191 & 3.30 \\ 
% Our model -B & \textbf{1.03} & \textbf{0.2131} & \textbf{2.09} \\ \hline
% Baseline 2 \cite{truong24b_interspeech} -L & 1.73 & 0.2297 & 2.47 \\ 
% Our model -L & \textbf{1.33} & \textbf{0.2213} & \textbf{2.05} \\ \hline
% \end{tabular}
% \end{table}

\section{Conclusion}
\label{sec:conclusion}

In this paper, we proposed a novel architecture utilised GR-KAN for anti-spoofing speech systems. Our model replaces fully-connected layer by GR-KAN, which is powerful in function approximation and dimensionality reduction. This placement can enhance the system’s continual learning capability when GR-KAN is positioned between the SSL model and the Conformer encoder as a feature projection layer. Our new model outperforms the baseline models and other competitive models on the ASVspoof 2021 evaluation set, with a relative reduction of EER of up to 54.0\% on LA set, and 67.1\% on DF set. Additionally, the ablation study proves the effectiveness of GR-KAN in continual learning with various SSL models of different sizes and architectures.

% \section{Acknowledgement}

% \ifinterspeechfinal
%      The Interspeech 2025 organisers
% \else
%      The authors
% \fi
% would like to thank ISCA and the organising committees of past Interspeech conferences for their help and for kindly providing the previous version of this template.

\bibliographystyle{IEEEtran}
\bibliography{mybib}

\end{document}